\documentclass[10pt,aps,pra,twocolumn,superscriptaddress]{revtex4-1} 

\usepackage{amsmath}
\usepackage{graphicx}
\usepackage{amsfonts}
\usepackage{color}
\usepackage{times}
\usepackage{amssymb}
\usepackage{mathrsfs}
\usepackage{hyperref}
\usepackage{textcomp}
\usepackage{multirow}
\usepackage{fancyhdr}

\newcommand{\ket}[1]{\left|#1\right>}
\newcommand{\bra}[1]{\left<#1\right|}

\newcommand{\eref}[1]{Eq.~(\ref{#1})}
\newcommand{\erefs}[1]{Eqs.~(\ref{#1})}

\newcommand{\dd}{\mathrm{d}} 
\newcommand{\ee}{\mathrm{e}} 
\newcommand{\ii}{\mathrm{i}}

\newcommand{\ha}{\hat{a}}
\newcommand{\had}{\hat{a}^\dagger}

\begin{document}

\title{Conditionally generating a mesoscopic superposition of N00N states}
\author{Falk T\"{o}ppel}
\email[]{falk.toeppel@mpl.mpg.de} 
\affiliation{Max Planck Institute for the Science of Light, G\"{u}nther-Scharowsky-Stra{\ss}e 1/Bldg. 24, 91058 Erlangen, Germany}
\affiliation{Institute for Optics, Information and Photonics, Universit\"{a}t Erlangen-N\"{u}rnberg, Staudtstra{\ss}e 7/B2, 91058 Erlangen, Germany}
\affiliation{Erlangen Graduate School in Advanced Optical Technologies (SAOT), Paul-Gordan-Stra{\ss}e 6, 91052 Erlangen, Germany}
\author{Maria V. Chekhova}
\affiliation{Max Planck Institute for the Science of Light, G\"{u}nther-Scharowsky-Stra{\ss}e 1/Bldg. 24, 91058 Erlangen, Germany}
\affiliation{Institute for Optics, Information and Photonics, Universit\"{a}t Erlangen-N\"{u}rnberg, Staudtstra{\ss}e 7/B2, 91058 Erlangen, Germany}
\author{Gerd Leuchs}
\affiliation{Max Planck Institute for the Science of Light, G\"{u}nther-Scharowsky-Stra{\ss}e 1/Bldg. 24, 91058 Erlangen, Germany}
\affiliation{Institute for Optics, Information and Photonics, Universit\"{a}t Erlangen-N\"{u}rnberg, Staudtstra{\ss}e 7/B2, 91058 Erlangen, Germany}
%


%
\begin{abstract}
\noindent
We study the generation of superpositions of N00N states by {overlapping} few-photon-subtracted squeezed vacuum states and coherent states on a 50/50 beam splitter. Assuming parameters that are feasible with current technology results {in output states with} mean photon numbers of several tens. We show that the generated quantum states violate local realism even in the presence of considerable loss. The occurrence of strong quantum correlations for mesoscopic quantum states of light is particularly beneficial for coupling light and matter quantum systems.
\end{abstract}
\maketitle


Entanglement at macroscopic scales is an utterly thought-provoking topic \cite{Schroedinger}. To explore the quantum-classical boundary, different mesoscopic and macroscopic quantum systems have been studied experimentally \cite{Julsgaard2001,Esteve2008,PhysRevLett.100.253601,PhysRevLett.109.150502}. Still, the emergence of classical laws like local realism for large physical systems remains debated \cite{RevModPhys.85.471}. An intriguing macroscopically entangled system in quantum optics is a N00N state, 
i.e. an equally weighted superposition of $N$ photons all being in one or the other of two modes. 
However, N00N states with large photon numbers are extremely challenging to produce in the lab. 
Here, we show that interfering a photon-subtracted squeezed vacuum state with a coherent state on a 50/50 beam splitter results in a quantum state that approximates a superposition of several large N00N states. We find that the generated state violates local realism. This combination of mesoscopic photon numbers and strong quantum correlations is especially appealing for quantum applications that require efficient light-matter interaction, e.g. in optomechanics and in quantum information science.  
The results attained may also be applied to other bosonic systems \cite{PhysRevLett.104.043601,1402-4896-2012-T147-014028}.

N00N states are primarily known for applications in quantum metrology \cite{PhysRevA.54.R4649,PhysRevA.55.2598}. Probabilistic methods to produce N00N states of arbitrary $N$ have been proposed \cite{PhysRevA.65.052104,PhysRevA.68.052315,PhysRevLett.99.163604}. In experiment, three, four and five-photon states have been generated using post-selection \cite{Mitchell2004,Walther2004,Afek2010}. 
A counterpart to N00N states in the continuous-variable regime are entangled coherent states, e.g.
\begin{align}
\label{eq:ECS}
\frac{1}{\sqrt{2(1\pm\ee^{-2|\lambda|^2})}}(\ket{\lambda}_1\!\ket{0}_2\pm\ket{0}_1\!\ket{\lambda}_2).
\end{align}
Therein, $\lambda$ denotes the coherent displacement. Entangled coherent states can be created from single-mode cat states \cite{PhysRevLett.71.2360,PhysRevA.80.022111,PhysRevA.88.052335,Ourjoumtsev09}.
However, it is difficult to generate cat states with high mean photon numbers \cite{Ourjoumtsev09}. 
Since \eref{eq:ECS} can be represented as superpositions of many N00N states, entangled coherent states have also been considered for metrology applications \cite{PhysRevA.64.063814,PhysRevLett.107.083601} and found to perform better than N00N states in the presence of dissipation \cite{PhysRevA.88.043832,PhysRevA.89.053812}.

For a superposition of N00N states it is for sure known that all photons are in one of two modes. However, neither is it knowable in which mode all the photons are nor is it predictable how many there are. This very quantum feature is of interest beyond metrology applications. For large photon numbers such superposition states are suitable to entangle two macroscopic objects, e.g. mirrors \cite{PhysRevA.92.012126}. Furthermore, entangled quantum states of light carrying many photons are appealing for coupling quantum light and atomic ensembles, e.g. in a quantum memory \cite{arXiv:1510.02665}.

%
\begin{figure*}[t]
\begin{center}
\includegraphics[width=0.9\textwidth]{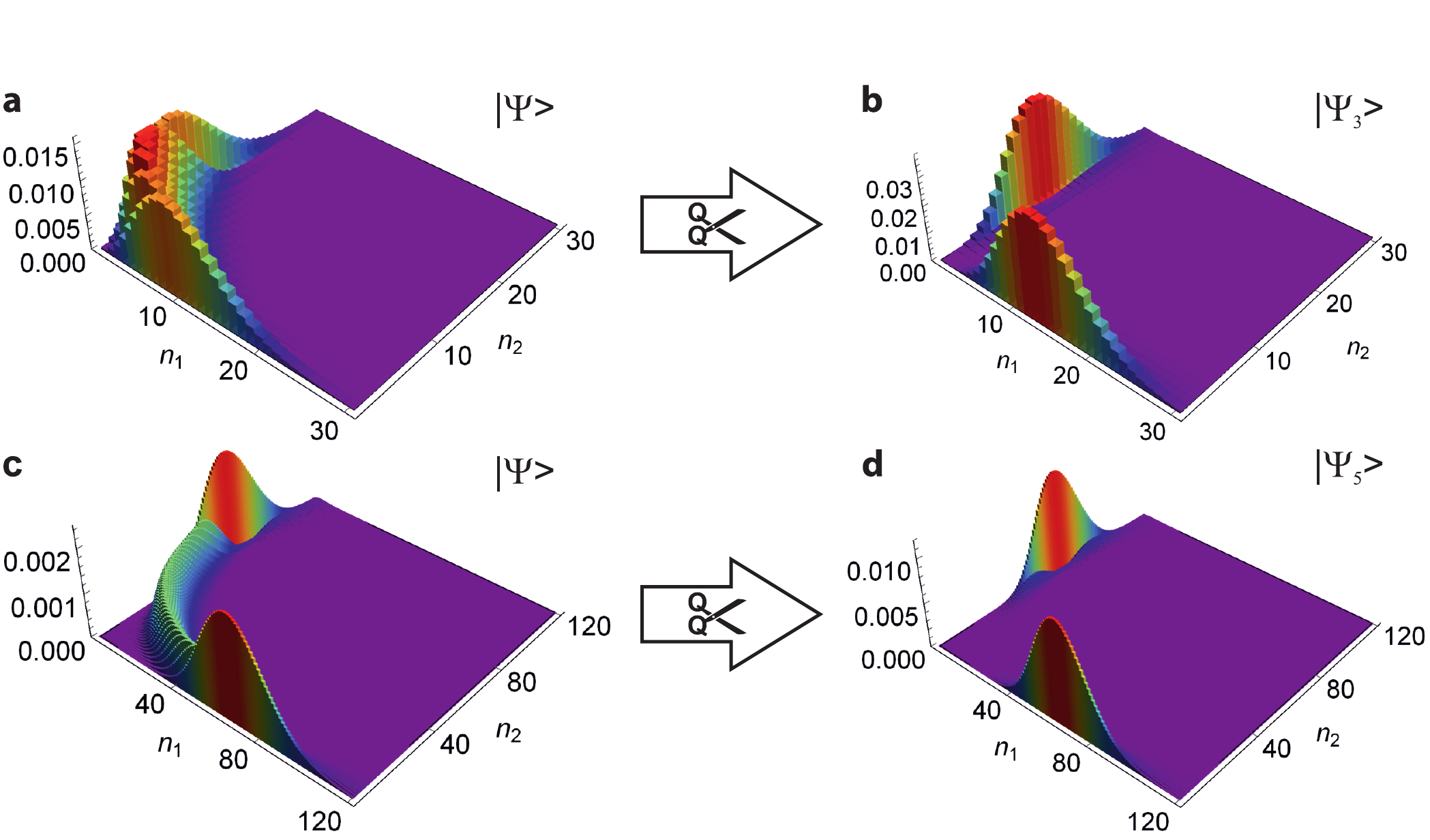}
\caption{\label{fig:pn_dist_input}Comparison of the photon number distributions with and without applying photon subtraction. \textbf{a}, \textbf{c}, Photon number distributions of a superposition of a squeezed state ($\zeta=1$ in \textbf{a} and $\zeta=2$ in \textbf{c}) and a coherent state ($\lambda=2.68\ii$ in \textbf{a} and $\lambda=5.81\ii$ in \textbf{c}) on a 50/50 beam splitter. See \eref{eq:input_state}. \textbf{b}, \textbf{d}, Using the setup shown in Fig.~\ref{fig:scheme} with $\tau=0.9$ and heralding on $m=3$ (\textbf{b}) or $m=5$ (\textbf{d}) clicks at the detector, we obtain a photon number distribution that {is a very good approximation to} the one of a superposition of N00N states. See \eref{eq:final_state}.}
\end{center}
\end{figure*}
%

In the experiment reported in Ref. \cite{Afek2010} a squeezed vacuum state $\ket{\zeta}$ and a coherent state $\ket{\lambda}$, with appropriately chosen squeezing parameter $\zeta$ and coherent displacement $\lambda$, have been superposed on a 50/50 beam splitter. The resulting quantum state takes the form: 
\begin{align}
\label{eq:input_state}
\ket{\Psi}_{12}=\hat{U}^\mathrm{BS}_{12}(1/2)\ket{\zeta}_1\!\ket{\lambda}_2.
\end{align}
Here, $\hat{U}^\mathrm{BS}_{ij}(\tau)=\exp[\arccos(\sqrt{\tau})(\had_i\ha_j-\had_j\ha_i)]$ is the unitary operator describing a loss-less beam splitter of transmittance $\tau$ and $\had_i$ ($\ha_i$) is the creation (annihilation) operator of the mode $i$.
The characteristic feature of a N00N state, called super-resolution, has been obtained after post-selecting on specific coincidence count events that add up to a total photon number $N=5$. The quantum state $\ket{\Psi}$ has further been examined for metrology applications in Refs.~\cite{PhysRevA.81.033819,1367-2630-13-8-083026}.

The state $\ket{\Psi}$ of \eref{eq:input_state} resembles a superposition of N00N states only for small mean photon numbers, i.e. small values of $|\zeta|$ and $|\lambda|$. Apart from two N00N-like `wings' one observes additional contributions to the photon number distribution when considering larger mean photon numbers. As an example thereof,  Fig.~\ref{fig:pn_dist_input}~a shows the photon number distribution of $\ket{\Psi}$ for a mean photon number of 9.33 ($\zeta=1$, $\lambda=2.68\ii$). In order to acquire a highly entangled mesoscopic quantum state of light it would be desirable to remove the non-N00N components from the state without destroying the superposition.

\begin{figure}[b]
\begin{center}
\includegraphics[width=0.35\textwidth]{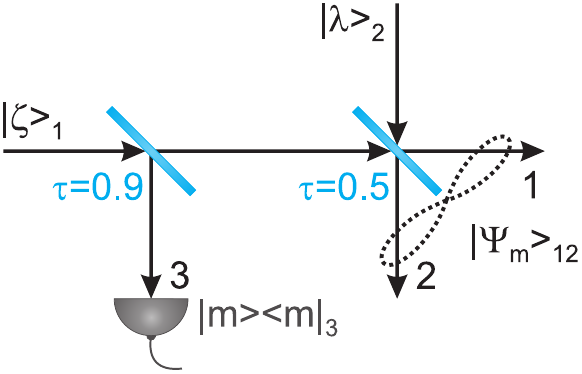}
\caption{\label{fig:scheme} Proposed experimental implementation. To generate the quantum state $\ket{\Psi_m}$ of \eref{eq:final_state}, one first subtracts by means of a highly transmitting beam splitters $m$ photons from a squeezed vacuum state $\ket{\zeta}$ and afterwards superposes it with a coherent state $\ket{\lambda}$ on a 50/50 beam splitter.}
\end{center}
\end{figure}

This can be achieved by subtracting photons from the squeezed vacuum before mixing it with the coherent state. When matching carefully the number of subtracted photons $m$ with the parameters $\zeta$ and $\lambda$, only the N00N-like 'wings' remain as is seen in Fig.~\ref{fig:pn_dist_input} b. The resulting quantum state closely approximates a superposition of N00N states. It contains on average some tens {of} photons when using realistic values for $m$, $\zeta$ and $\lambda$. A possible experimental implementation is depicted schematically in Fig.~\ref{fig:scheme}. Before superposing the squeezed vacuum state with the coherent state on a 50/50 beam splitter, the squeezed light impinges on a tapping beam splitter of high transmittance $\tau$. We obtain the output state $\ket{\Psi_m}_{12}$ occupying modes 1 and 2 if the photon number resolving detector (PNRD) in mode 3 registers exactly $m$ photons:
\begin{align}
\label{eq:final_state}
\ket{\Psi_m}_{12}=\frac{1}{\sqrt{\mathcal{N}}}\hat{U}^\mathrm{BS}_{12}(1/2)\bra{m}_3\hat{U}^\mathrm{BS}_{13}(\tau)\ket{\zeta}_1\!\ket{\lambda}_2\!\ket{0}_3.
\end{align}
Therein, $\mathcal{N}$ ensures a proper normalization and quantifies the heralding rate. PNRDs that resolve a few photons and have a high detection efficiency are currently available \cite{Rosfjord:06,Calkins:13}. In experiment the subtraction of three photons from squeezed vacuum has been demonstrated \cite{PhysRevA.82.031802}. 

%
%

%
\begin{figure*}[t!]
\begin{minipage}{0.53\textwidth}
\!\!\!\!\!\!\includegraphics[width=\columnwidth]{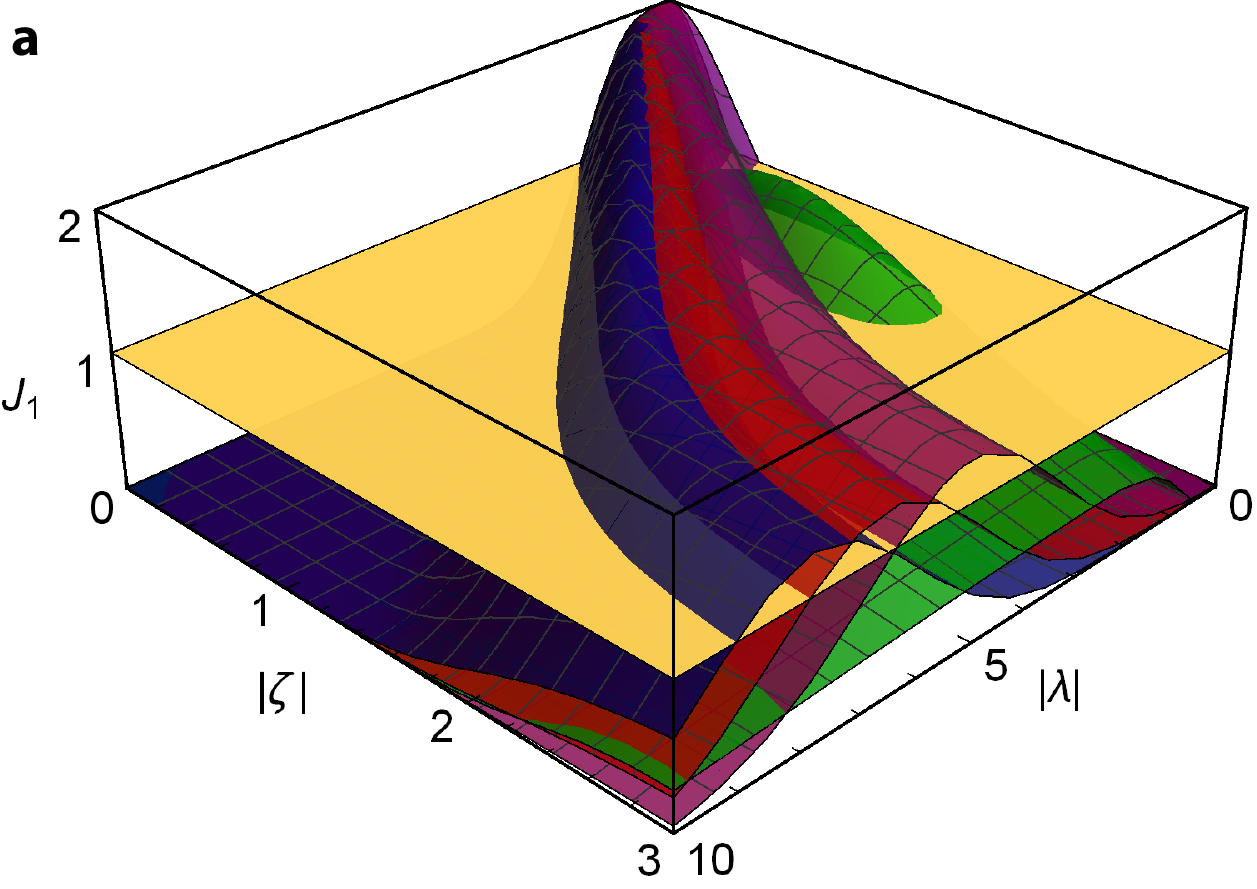}
\end{minipage}
\begin{minipage}{0.05\textwidth}
\includegraphics[width=\columnwidth]{white}
\end{minipage}
\begin{minipage}{0.4\textwidth}
\includegraphics[width=\columnwidth]{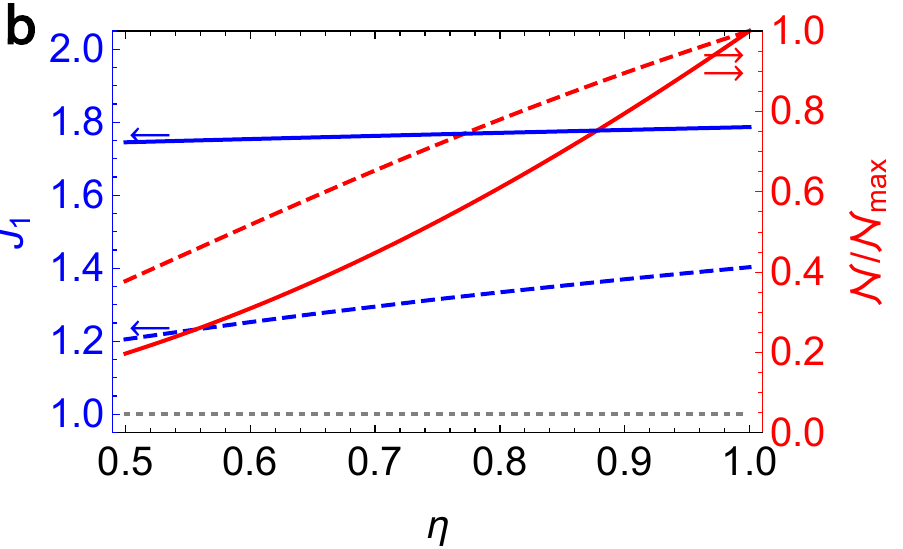}
\\[3mm]
\includegraphics[width=\columnwidth]{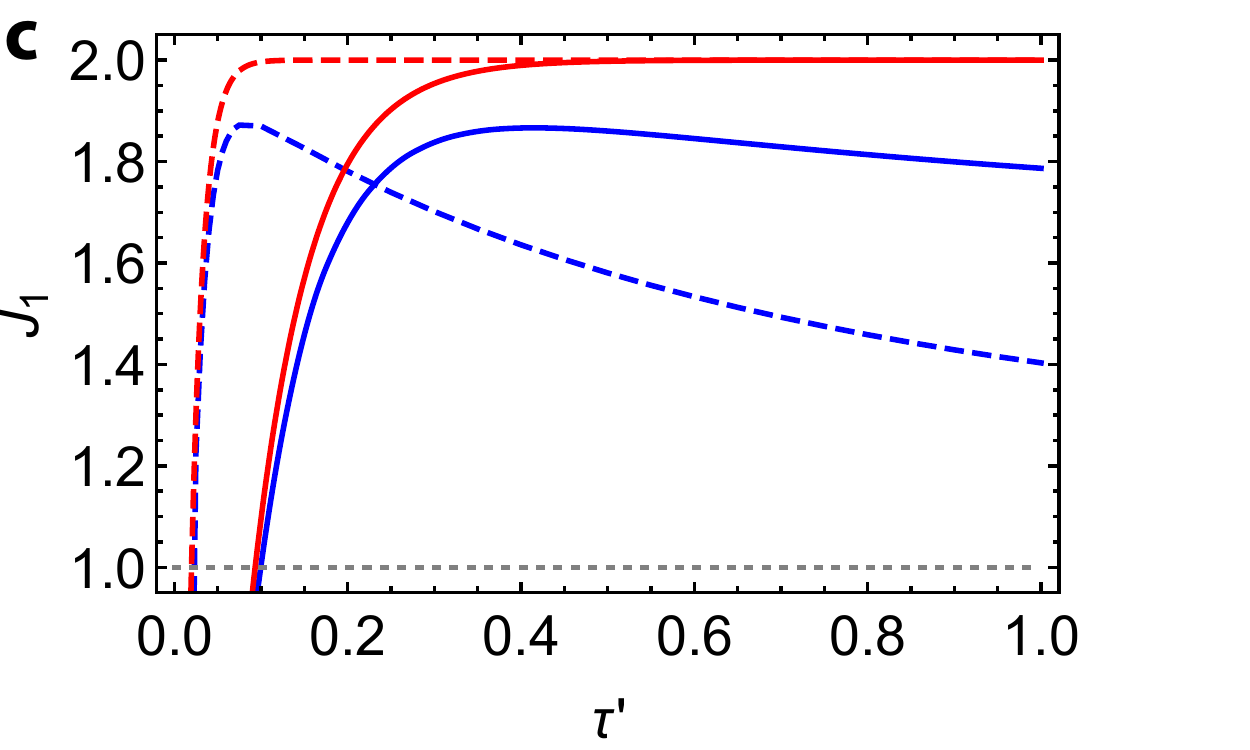}
\end{minipage}
\caption{\label{fig:bell_violation} Violation of the Bell inequality (\ref{eq:Bell_J1}). \textbf{a}, Maximal value $J_1$ of \eref{eq:Bell_J1} for the state $\ket{\Psi_m}$, defined in \eref{eq:final_state}, using different values of $\zeta=|\zeta|$, $\lambda=\ii|\lambda|$ and $m$. The yellow plane represents $J_1=1$. For $m=0$ (green), i.e. the state $\ket{\Psi}$ of \eref{eq:input_state}, the Bell inequality (\ref{eq:Bell_J1}) is only violated for small values of $|\zeta|$ and $|\lambda|$. In contrast, for $m=3$ (purple), $m=5$ (red) and $m=7$ (blue) the Bell inequality is violated for much larger values of $|\zeta|$ and $|\lambda|$, i.e. larger mean photon numbers. In addition, the violation is stronger than for $m=0$. 
\textbf{b}, Bell observable $J_1$ (blue) 
evaluated for the heralded output state $\ket{\Psi_m}$ (solid line: $\zeta=1$, $m=3$; dashed line: $\zeta=2$, $m=5$) with an optimized coherent displacement plotted as a function of the detection efficiency $\eta$ of the PNRDs, used for heralding. Inefficient detection only moderately diminishes the Bell violation $J_1>1$. However, the heralding rate $\mathcal{N}$ (red) drops significantly with decreasing $\eta$. \textbf{c}, Impact of the transmittance $\tau'$ of output modes 1 and 2 on the maximal Bell violation (solid blue line: $\zeta=1$, $m=3$; dashed blue line: $\zeta=2$, $m=5$). 
For comparison the Bell violation for the entangled coherent state of \eref{eq:ECS} with the same mean photon number is shown in red.
}
\end{figure*}
%

To quantify how well the quantum state under consideration resembles a superposition of N00N states, we evaluate its overlap with a superposition of the form
\begin{align}
\label{eq:superposition_n00n}
|\Psi_m'\rangle_{12}&=\ii^m\sqrt{\mathcal{P}_0}\ket{0}_1\!\ket{0}_2\\
&\quad+\sum_{N=1}^\infty\ii^{N+m}\sqrt{\frac{\mathcal{P}_N}{2}}\bigl[\ket{N}_1\!\ket{0}_2+(-1)^m\ket{0}_1\!\ket{N}_2\bigr],\nonumber
\end{align}
where
\begin{align}
\label{eq:marginal_distribution}
\mathcal{P}_N=|\langle\Psi_m|N\rangle\!\ket{0}|^2/\sum_{n=0}^\infty|\langle\Psi_m|N\rangle\!\ket{0}|^2.
\end{align}
{We remark that $\mathcal{P}_N$ has the shape of the peaks along the edges of the photon number distributions shown in Fig.~\ref{fig:pn_dist_input} b and d.}
The fidelity $\mathcal{F}=|\langle\Psi_m'|\Psi_m\rangle|$ depends for given $m$ on the parameters $\zeta$ and $\lambda$ of the input state. We find for the states depicted in Figs.~\ref{fig:pn_dist_input}~b ($\zeta=1$, $\lambda=2.82\ii$, $m=3$) and d ($\zeta=2$, $\lambda=6.1\ii$, $m=5$) an overlap of $\mathcal{F}=0.945$ and $\mathcal{F}=0.840$, respectively, with the corresponding superposition of N00N states $|\Psi_m'\rangle$. In order to give a glance beyond the particular examples considered so far, Tab.~\ref{tab:comparison} summarizes for different parameter settings the maximal fidelity, the rate of heralding the state $\ket{\Psi_m}$ as well as the mean value and the standard deviation of the photon number distribution of the N00N state superposition $\ket{\Psi_m'}$. The distribution $\mathcal{P}_N$, defined in \eref{eq:marginal_distribution}, is not Poissonian, i.e. $\ket{\Psi_m'}$ is different from an entangled coherent state. In fact, the distribution is mostly broader as we can infer from the last two columns of Tab.~\ref{tab:comparison}. Only for small values of $|\zeta|$ and $|\lambda|$ $\mathcal{P}_N$ is narrower than a Poissonian distribution. 

{The N00N-like features in the photon number distribution even withstand experimental imperfections as is shown in Appendix~\ref{appx:figures}.} {There, inefficient photon subtraction, transmission losses, improper phase choice and multi-mode input states are considered.}

{A} photon subtracted squeezed vacuum {state resembles a} single-mode cat {state} \cite{PhysRevA.55.3184}. Hence, {it is not surprising that} its superposition with {a} coherent {state yields} a quantum state similar to an entangled coherent state. {However}, we find the fidelity $\mathcal{F}$ to be even larger than the overlap between the photon subtracted-squeezed vacuum state and its corresponding single-mode cat state.

N00N states are known to be highly entangled. There exist Bell inequalities, formulated in terms of the Q-function (see Appendix~\ref{appx:q-function}), that show violation of local realism for any $N$ \cite{PhysRevA.76.052101}. To verify that the heralded state $\ket{\Psi_m}$ is entangled and indeed violates local realism, we consider the following Bell inequality given in equation~(22) of Ref.~\cite{PhysRevA.76.052101}:
\begin{align}
\label{eq:Bell_J1}
J_1&=\mathcal{Q}(\alpha)+\mathcal{Q}(\beta)+\mathcal{Q}(\gamma)+\mathcal{Q}(\delta)-\mathcal{Q}(\alpha,\beta)-\mathcal{Q}(\alpha,\gamma)\nonumber\\
&\quad-\mathcal{Q}(\alpha,\delta)-\mathcal{Q}(\beta,\gamma)-\mathcal{Q}(\beta,\delta)-\mathcal{Q}(\gamma,\delta)\leq 1.
\end{align}
Therein, the quantities $\mathcal{Q}(\alpha_1,\alpha_2)=\pi^2 Q(\alpha_1,\alpha_2)$ and $\mathcal{Q}(\alpha_1)=\pi\int\dd^2 \alpha_2\,Q(\alpha_1,\alpha_2)$ are proportional to the two-mode and single-mode Q-function{s} of the state under consideration. 
We remark that the Bell inequality (\ref{eq:Bell_J1}) can be tested with simple on/off-detectors \cite{PhysRevLett.82.2009,PhysRevA.76.052101}.

We examine the Bell inequality (\ref{eq:Bell_J1}) for the state $\ket{\Psi_m}$ of \eref{eq:final_state} by inserting the corresponding Q-function provided in \eref{eq:q_func_out} of Appendix~\ref{appx:q-function}. Figure~\ref{fig:bell_violation} a shows the maximal value of $J_1$ for different values of $\zeta$, $\lambda$ and $m$. In general, we conclude that the violation of the Bell inequality (\ref{eq:Bell_J1}) is stronger for the heralded output state $\ket{\Psi_m}$ than it is for the state $\ket{\Psi}$, given in \eref{eq:input_state}. Moreover, we find that with photon subtraction the Bell inequality $J_1<1$ is violated for much larger values of $|\zeta|$ and $|\lambda|$, i.e. for larger mean photon numbers, than without it.  
The amount of Bell violation {for fixed $\zeta$} degrades only very little with increasing $m$. 
This observation and further calculations reveal that the generated quantum state $\ket{\Psi_m}$ violates local realism also for mean photon numbers of a few hundred. In the limit of $|\zeta|\rightarrow0$ and $|\lambda|\rightarrow0$ we find $J_1$ to approach 2 and strongly violating the Bell inequality (\ref{eq:Bell_J1}), although in this limit we consider two vacuum states as input. This seemingly unphysical result is an artifact since the probability $\mathcal{N}$ to successfully generate $\ket{\Psi_m}$ also approaches 0 for $|\zeta|\rightarrow0$ and $|\lambda|\rightarrow0$. Henceforth, the state preparation is never successful for $\zeta=0$ and $\lambda=0$.

Next, we study some possible experimental imperfections that might affect the quality of the state preparation.
Figure~\ref{fig:bell_violation} b shows how the amount of violation of the Bell inequality (\ref{eq:Bell_J1}) changes with the detection efficiency $\eta$ of the PNRD {used for photon subtraction}. The change is very little for the case $\zeta=1$, $m=3$ and an optimized coherent displacement $\lambda_\mathrm{opt}$. Nevertheless, the detection efficiency has an impact on the rate to successfully herald the output state, which is also depicted in Fig.~\ref{fig:bell_violation} b. For the parameter set $\zeta=2$, $\lambda_\mathrm{opt}$ and $m=5$ the violation is decreasing slightly faster, but still the Bell inequality (\ref{eq:Bell_J1}) is clearly violated even for small detection efficiencies $\eta$.  
For mean photon numbers of a few hundred
, the critical detection efficiency needed to observe $J_1>1$, is on the order of $60\%$ to $70\%$.  

The effect of transmission losses in modes 1 and 2 after a successful state generation on the violation of the Bell inequality (\ref{eq:Bell_J1}) is analyzed in Fig.~\ref{fig:bell_violation} c. 
The quantum state $\ket{\Psi_m}$ consists of no or few photons in one mode and many photons in the other. With growing transmission losses, it is more likely that the mode containing only a few photons is emptied completely. Therefore, the lossy quantum state becomes more N00N-like for moderate loss (see Fig.~7 in Appendix \ref{appx:figures}). For low transmission coefficients $\tau'$ however, the vacuum component is dominant and no violation of the Bell inequality is to observe.
We conclude that the entanglement of the quantum state $\ket{\Psi_m}$ survives in the presence of small experimental imperfections. The Q-function that we have used to define $J_1$ in the case of inefficient detection and transmission losses is provided in \erefs{eq:q_func_out_lossy} and (\ref{eq:q_func_out_transmission}) of Appendix \ref{appx:losses}. 

Recently, Ref.~\cite{Birrittella:14} examined the phase-shift measurement sensitivity for small phase shifts when using a photon number parity measurement in one output of a Mach-Zehnder interferometer. It was found that, for given squeezing and coherent displacement, mixing of a photon-subtracted squeezed vacuum state and a coherent state at the input beam splitter yields a higher sensitivity than without photon subtraction. Interestingly, the results do not depend on how well the superposition of the two input states resembles a superposition of N00N states. For measurement {schemes not relying} on parity measurements this might be different. Furthermore, the performance of the quantum state $\ket{\Psi_m}$ for $m=1$ in measuring nonlinear phase shifts was analyzed in Ref.~\cite{PhysRevA.86.043828}.

The proposed scheme is a method to create novel entangled macroscopic quantum states of light. In contrast to Ref.~\cite{PhysRevLett.100.253601} and Refs.~\cite{Bruno13,Lvovsky13} using squeezing and coherent displacement, respectively, to amplify entangled quantum states of a few photons to macroscopic scales, here we exploit {the interference of two multi-photon quantum states} at a beam splitter. 

As an extension of the setup displayed in Fig.~\ref{fig:scheme}, we may consider two orthogonal polarization modes in each of the input modes 1 and 2. When these polarization modes are excited by squeezed vacuum states and coherent states, respectively, it is possible to subtract a polarization qubit. 
This is achieved by means of a half-wave plate and a polarizing beam splitter whose both output ports are monitored by PNRDs. When only one of these two PNRDs registers a photon in mode 3, the qubit state $\cos\theta\ket{1_h,0_v}_3+\sin\theta\ket{0_h,1_v}_3$ has been subtracted and the mesoscopic qubit $\cos\theta\ket{\Psi_1}_h\ket{\Psi}_v+\sin\theta\ket{\Psi}_h\ket{\Psi_1}_v$ is established in the output modes 1 and 2. Note that the angle $\theta$ is determined by the orientation of the half-wave plate.

%
\begin{table}[h]
\begin{ruledtabular}
\begin{tabular}{ccccccc}
$m$ & $\zeta$ & $\mathcal{F}$ & $\lambda_\mathrm{opt}$ & $\mathcal{N}$ & ${\mathrm{E}(N)}$ & ${\sqrt{\mathrm{Var}(N)}}$ \\\hline
$1$ & $0.5$ & $0.996$ & $1.18\ii$ & $2.27\cdot10^{-2}$ & $2.98$ & $1.65$ \\
~ & $1.0$ & $0.961$ & $1.89\ii$ & $8.76\cdot10^{-2}$ & $6.47$ & $3.13$ \\
~ & $1.5$ & $0.903$ & $2.36\ii$ & $1.61\cdot10^{-1}$ & $11.2$ & $4.68$ \\\hline
$2$ & $0.5$ & $0.981$ & $1.26\ii$ & $2.05\cdot10^{-3}$ & $2.79$ & $2.20$ \\
~ & $1.0$ & $0.943$ & $2.22\ii$ & $1.78\cdot10^{-2}$ & $9.91$ & $4.13$ \\
~ & $1.5$ & $0.887$ & $3.13\ii$ & $6.18\cdot10^{-2}$ & $19.5$ & $6.28$ \\\hline
$3$ & $0.5$ & $0.991$ & $1.60\ii$ & $1.18\cdot10^{-4}$ & $5.17$ & $2.47$ \\
~ & $1.0$ & $0.945$ & $2.68\ii$ & $3.56\cdot10^{-3}$ & $14.4$ & $4.92$ \\
~ & $1.5$ & $0.885$ & $3.75\ii$ & $2.52\cdot10^{-2}$ & $28.2$ & $7.57$ \\\hline
$4$ & $1.0$ & $0.943$ & $3.05\ii$ & $7.60\cdot10^{-4}$ & $18.7$ & $5.62$ \\
~ & $1.5$ & $0.884$ & $4.29\ii$ & $1.08\cdot10^{-2}$ & $36.9$ & $8.67$ \\
~ & $2.0$ & $0.838$ & $5.21\ii$ & $3.95\cdot10^{-2}$ & $54.5$ & $11.0$ \\\hline
$5$ & $1.0$ & $0.943$ & $3.57\ii$ & $1.30\cdot10^{-6}$ & $23.0$ & $6.25$ \\
~ & $1.5$ & $0.883$ & $4.77\ii$ & $4.75\cdot10^{-3}$ & $45.7$ & $9.64$ \\
~ & $2.0$ & $0.838$ & $5.81\ii$ & $2.59\cdot10^{-2}$ & $67.6$ & $12.3$ \\\hline
$7$ & $1.0$ & $0.942$ & $3.97\ii$ & $8.30\cdot10^{-6}$ & $31.6$ & $7.33$ \\
~ & $1.5$ & $0.882$ & $5.62\ii$ & $9.65\cdot10^{-4}$ & $63.3$ & $11.4$ \\
~ & $2.0$ & $0.837$ & $6.84\ii$ & $1.17\cdot10^{-2}$ & $93.7$ & $14.5$ \\
\end{tabular}
\end{ruledtabular}
\caption{\label{tab:comparison}Results for different parameter settings. In the first three columns different values of the squeezing parameter $\zeta$ and the number of photons $m$, used for heralding, are given. The third column displays the maximal fidelity $\mathcal{F}=|\langle\Psi_m'|\Psi_m\rangle|$ after optimizing over the coherent displacement. The corresponding optimal coherent displacement $\lambda_\mathrm{opt}$ is given in the fourth column. The fifth column shows the heralding rate of the state $\ket{\Psi_m}$, i.e. the normalization constant $\mathcal{N}$ occurring in \eref{eq:final_state}. The last two columns yield the mean value and the standard deviation of the distribution $\mathcal{P}_N$, defined in \eref{eq:marginal_distribution}, that is used to define the quantum state $\ket{\Psi_m'}$ via \eref{eq:superposition_n00n}.}
\end{table}
%

In summary, we have {proposed} a feasible experimental method, depicted in Fig.~\ref{fig:scheme}, to create a quantum state, given in \eref{eq:final_state}, that closely approximates a superpositions of large N00N states (see Fig.~\ref{fig:pn_dist_input} b and d). The method is based on photon subtraction, which can be implemented with state-of-the-art technology. The quantum states generated are strongly entangled as we have shown by violating the Bell inequality (\ref{eq:Bell_J1}) for a range of parameters $\lambda$ and $\zeta$ also in the presence of considerable experimental imperfections (see Fig.~\ref{fig:bell_violation}). The mean photon numbers of the states violating local realism can reach some tens for realistic values of the parameters (see Tab.~\ref{tab:comparison}). Quantum applications that require a highly efficient interaction between light and matter, e.g. in quantum information science and in optomechanics, may benefit from the outlined scheme to generate a strongly entangled quantum state with a mesoscopic mean photon number. Further technical advances, in particular allowing for higher squeezing, would even render mean photon numbers of a few hundred possible. Therefore, the proposed method is also an approach towards realizing an intriguing entangled macroscopic quantum state of light in experiment.

\appendix

%

\section{Deriving the Q-function of the output state $\ket{\Psi_m}$}
\label{appx:q-function}

In general, the Q-function of a $k$-mode density operator $\hat{\rho}$ is defined as 
\begin{align*}
Q(\alpha_1,\dots,\alpha_k;\hat{\rho})&=\frac{1}{\pi^k}\mathrm{tr}\Bigl[\hat{\rho}\ket{\alpha_1}\!\bra{\alpha_1}_1\!\otimes\!\dots\otimes\ket{\alpha_k}\!\bra{\alpha_k}_k\Bigr].
\end{align*}
The Q-function of the squeezed vacuum state after $m$ photons have been subtracted by means of a beam splitter with transmittance $\tau$ and a PNRD, in particular, takes the form:
\begin{align}
\label{eq:q_func_subtracted_SV}
Q(\alpha_1;\ket{\zeta_m})
&=\frac{\pi}{\mathcal{N}}\!\!\int\!\dd^2\alpha_3\, Q(\alpha_1';\ket{0,\zeta})Q(\alpha_3';\ket{0,0})\nonumber\\
&\quad\times P(\alpha_3;\ket{m}),
\end{align}
with 
\begin{align*}
\alpha_1'&=\sqrt{\tau}\alpha_1-\sqrt{1-\tau}\alpha_3\quad\mathrm{and}\quad\alpha_3'=\sqrt{1-\tau}\alpha_1+\sqrt{\tau}\alpha_3. 
\end{align*}
These substitutions are due to the action of the highly transmitting beam splitter in the scheme (see Fig.~\ref{fig:scheme}). In \eref{eq:q_func_subtracted_SV} $P(\alpha;\ket{m})$ denotes the P-function of the photon number state $\ket{m}$: 
\begin{align*}
P(\alpha;\ket{m})=\sum_{k=0}^n\binom{n}{k}\frac{1}{k!}\frac{\partial^k}{\partial\alpha}\frac{\partial^k}{\partial\alpha^*}\delta(\alpha),
\end{align*}
and $Q(\alpha;\ket{\lambda,\zeta})$ denotes the single-mode Q-function of the squeezed coherent state $\ket{\lambda,\zeta}$:
\begin{align*}
Q(\alpha;\ket{\lambda,\zeta})=\frac{1}{\pi\cosh|\zeta|}\exp\Bigl[-2\, \mathbf{c}\cdot \mathsf{V}^{-1}(\zeta)\cdot\mathbf{c}^\mathsf{T}\Bigr].
\end{align*}
Therein, we have defined the vector $\bf{c}=[\mathrm{Re}(\alpha-\lambda),\mathrm{Im}(\alpha-\lambda)]$, the correlation matrix
\begin{align*}
\mathsf{V}(\zeta)&=\mathsf{R}[\mathrm{arg}(\zeta)/2]
\cdot
\begin{bmatrix}
 \ee^{-2|\zeta|}+1 & 0\\
 0 & \ee^{2|\zeta|}+1 
\end{bmatrix}
\cdot
\mathsf{R}^\mathsf{T}[\mathrm{arg}(\zeta)/2],
\end{align*}
and the rotation matrix
\begin{align*}
\mathsf{R}(\varphi)=\begin{bmatrix}
 \cos\varphi & -\sin\varphi\\
 \sin\varphi & \cos\varphi
\end{bmatrix}.
\end{align*}
Considering the action of the 50/50 beam splitter in the setup (see Fig.~\ref{fig:scheme}), we finally arrive at the following Q-function for the state $\ket{\Psi_m}$, defined in \eref{eq:final_state}:
\begin{align}
\label{eq:q_func_out}
Q(\alpha_1,\alpha_2;\ket{\Psi_m})
&=Q((\alpha_1-\alpha_2)/\sqrt{2};\ket{\zeta_m})\nonumber\\
&\quad\times Q((\alpha_1+\alpha_2)/\sqrt{2};\ket{\lambda,0}).
\end{align}

\newpage

\section{Q-function in the presence of experimental imperfections}
\label{appx:losses}
The probability that in the scheme of Fig.~\ref{fig:scheme} exactly $m'$ photons are reflected into mode 3 is{,} according to \eref{eq:q_func_subtracted_SV}{,} given by
\begin{align}
\label{eq:normalization}
\mathcal{N}_{m'}\!&=\pi\!\!\int\!\dd^2\alpha_1\,\dd^2\alpha_3\, Q(\alpha_1';\ket{\zeta,0})\,Q(\alpha_3';\ket{0,0})\nonumber\\
&\quad\times P(\alpha_3;\ket{m'}).
\end{align}
This quantity ensures the normalization in \erefs{eq:final_state} and (\ref{eq:q_func_out}).   
The probability that a realistic PNRD with detection efficiency $\eta$ registers exactly $m$ out of $m'$ impinging photons {has a binomial distribution}. Therefore, when using a realistic PNRD for heralding, the Q-function of the output state is given by
\begin{align}
\label{eq:q_func_out_lossy}
&\frac{\pi^2}{\mathcal{N}_\eta}\!\sum_{m'=m}^\infty\!\!\binom{m'}{m}\eta^{m}(1-\eta)^{m'-m}\mathcal{N}_{m'}\, Q(\alpha_1,\alpha_2;\ket{\Psi_{m'}}),
\end{align}
with
\begin{align*}
\mathcal{N}_\eta=\sum_{m'=m}^\infty\!\!\binom{m'}{m}\eta^{m}(1-\eta)^{m'-m}\mathcal{N}_{m'},
\end{align*}
and $Q(\alpha_1,\alpha_2;\ket{\Psi_m})$ as given in \eref{eq:q_func_out}.

The Q-function that takes into consideration the transmission losses after a successful state preparation is obtained as
\begin{align}
\label{eq:q_func_out_transmission}
&\int\!\!\dd^2\beta_1\,\dd^2\beta_2\,Q(\sqrt{1-\tau'}\alpha_1+\sqrt{\tau'}\beta_1;\ket{0,0})\nonumber\\
&\times Q(\sqrt{1-\tau'}\alpha_2+\sqrt{\tau'}\beta_2;\ket{0,0})\nonumber\\
&\times Q(\sqrt{\tau'}\alpha_1-\sqrt{1-\tau'}\beta_1,\sqrt{\tau'}\alpha_2-\sqrt{1-\tau'}\beta_2;m,\ket{\Psi_m}).
\end{align}
To mimic transmission losses we have considered a hypothetical beam splitter of transmittivity $\tau'$ located in each of the output modes 1 and 2. 

%
\begin{figure}[h]
\begin{center}
\includegraphics[width=0.4\textwidth]{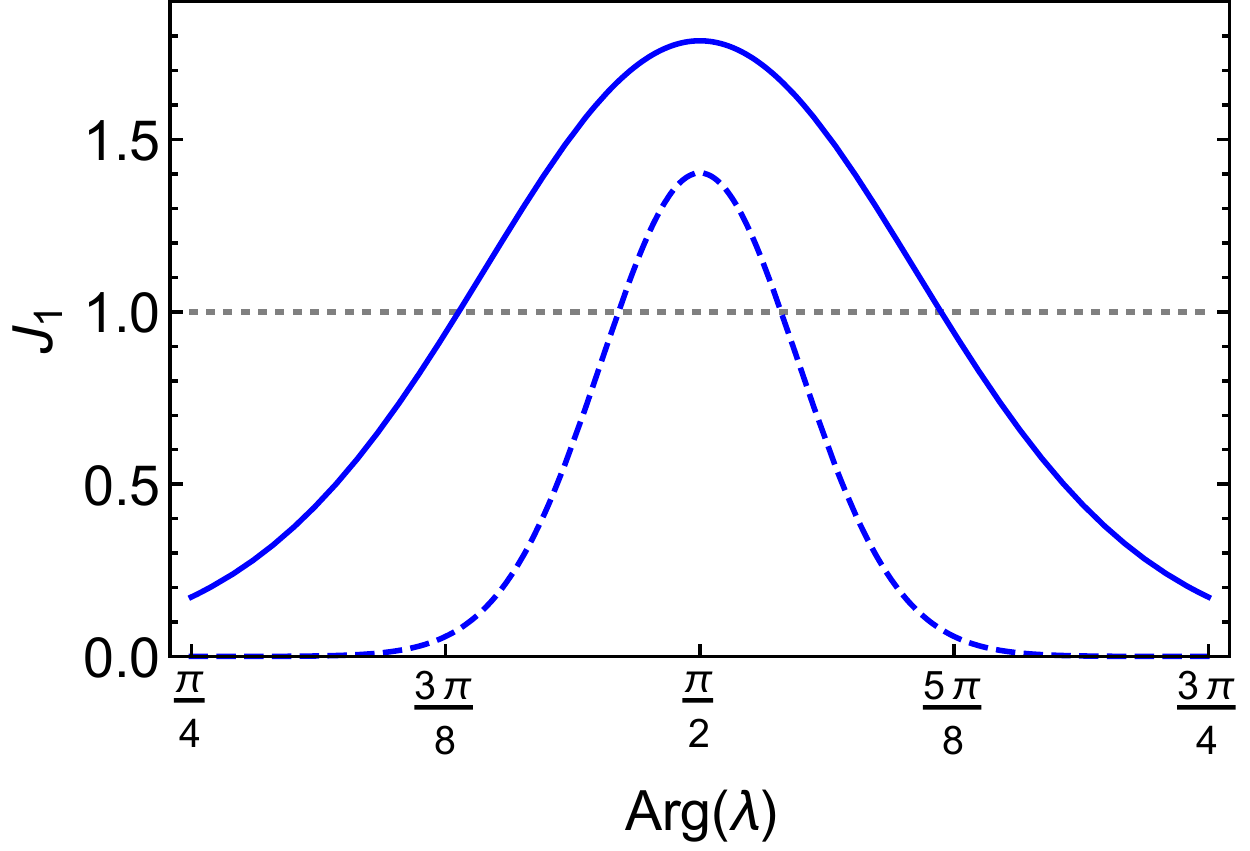}
\caption{\label{fig:bell_improper_phase} {Violation of the Bell inequality (\ref{eq:Bell_J1}) for improper phase of the coherent displacement $\lambda$. Solid line: $\zeta=1$, $|\lambda|=2.68$, $m=3$; dashed line: $\zeta=2$, $|\lambda|=5.81$, $m=5$.}}
\end{center}
\end{figure}
%

{The Q-functions of \erefs{eq:q_func_out_lossy} and \ref{eq:q_func_out_transmission} were used to study a violation of the Bell inequality (\ref{eq:Bell_J1}) for non-unit detection efficiency $\eta$ (see Fig.~\ref{fig:bell_violation} b) and in the presence of transmission losses $\tau'$ after a successful state generation (see Fig.~\ref{fig:bell_violation} c), respectively.} 

{The state preparation crucially depends on the relative phase difference between squeezing parameter $\zeta$ and coherent displacement $\lambda$. To quantify this phase sensitivity, Fig.~\ref{fig:bell_improper_phase} displays the Bell violation for a real-valued squeezing parameter $\zeta$ and different phases of the coherent displacement $\lambda$.} {One can see that the phase range for which the Bell inequality is still violated is rather broad, between $\pi/8$ and $\pi/4$.}

\section{Photon number distributions in the presence of experimental imperfections}
\label{appx:figures}

\begin{figure}[h]
\begin{center}
\includegraphics[width=0.24\textwidth]{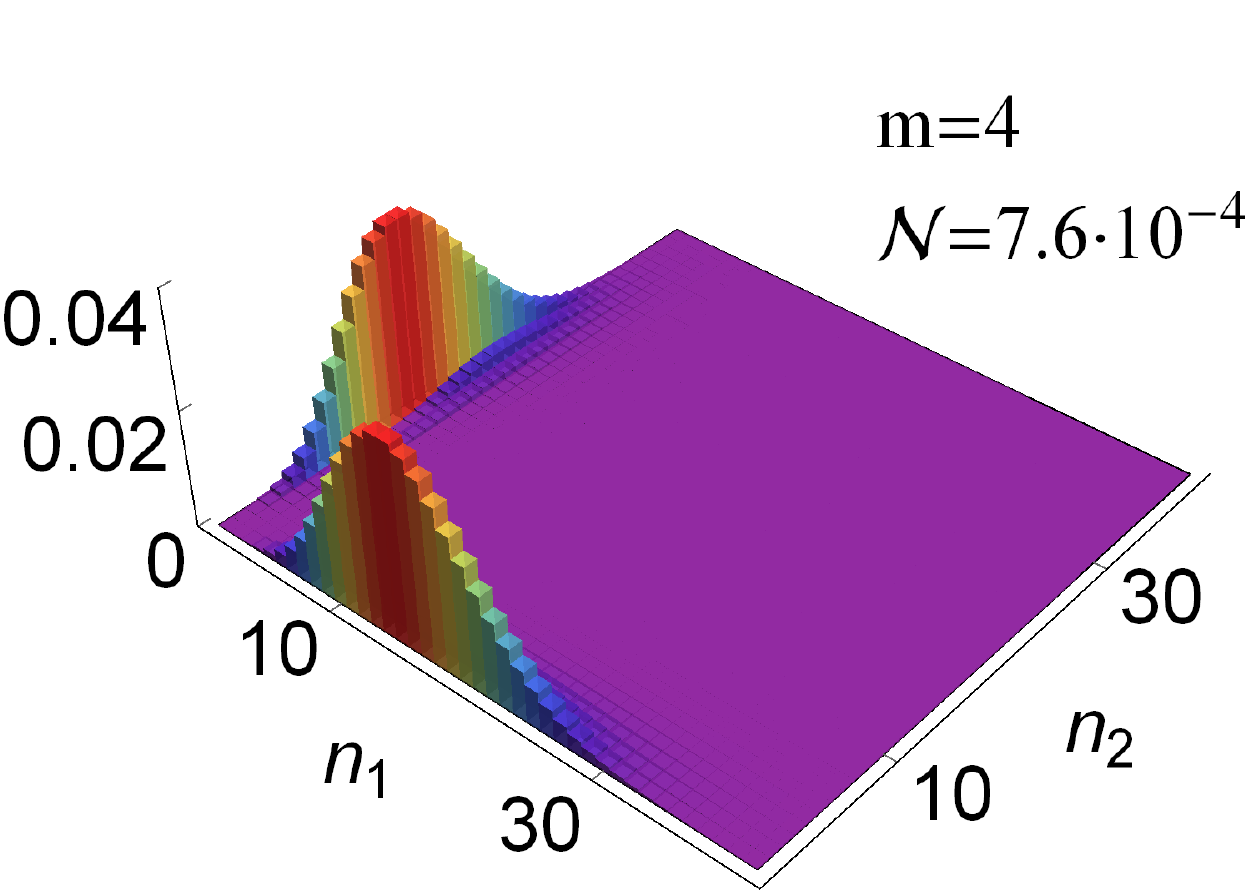}\;\;\includegraphics[width=0.24\textwidth]{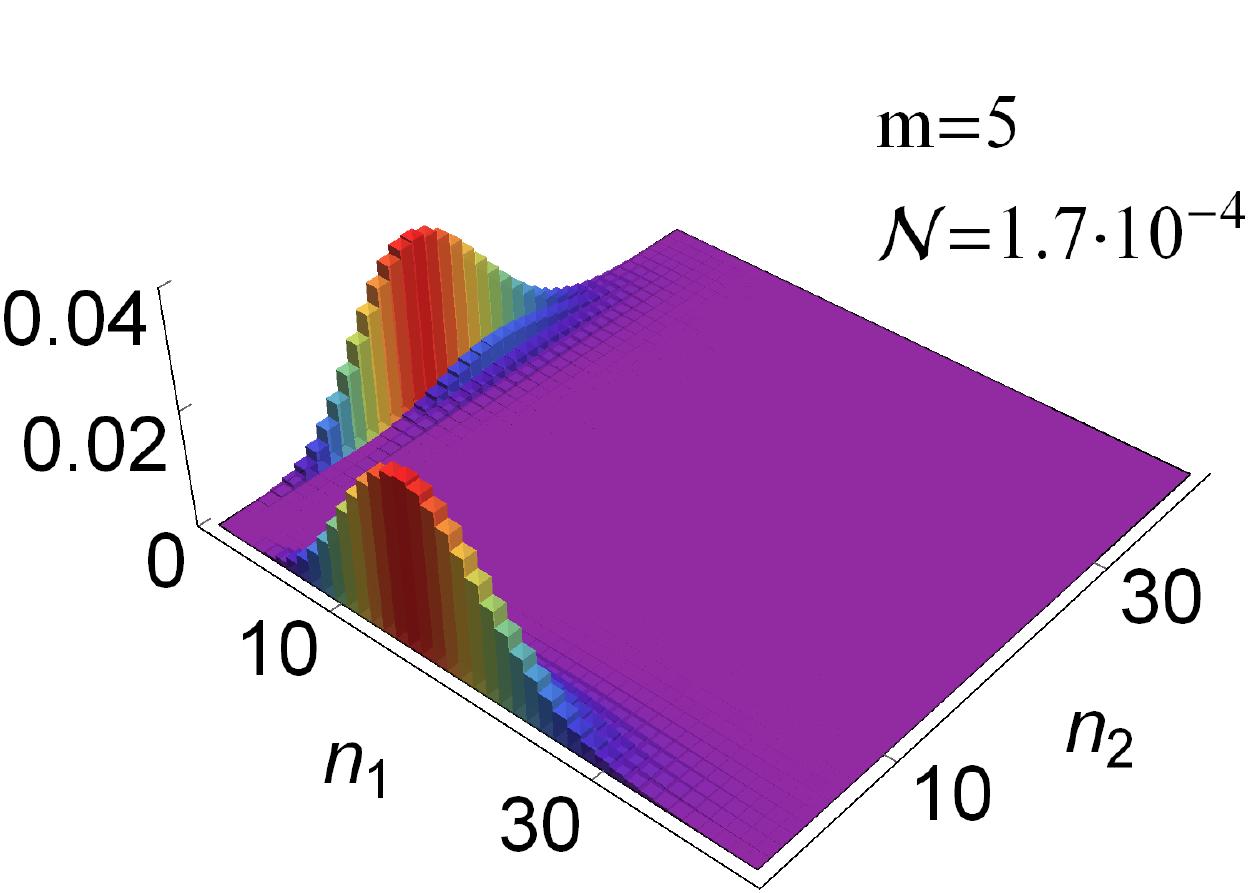}
\caption{\label{fig:pn_dist_lossy}Photon number distributions after superposing a photon-subtracted squeezed vacuum state ($\zeta=1$) and a coherent state ($\lambda=2.68\ii$) on a 50/50 beam splitter using the setup shown in Fig.~2 with $\tau=0.9$ and heralding on $m$ clicks at an ideal PNRD.}
\end{center}
\end{figure}

In Fig.~\ref{fig:pn_dist_lossy} the photon number distributions of $\ket{\Psi_m}$ for $\zeta=1$, $\lambda=2.68\ii$ (see Fig. 1 b) is plotted for different numbers of subtracted photons $m$. We find that the N00N-like 'wings' are also present for a slightly imperfect preparation. {This observation coincides with the robustness of the Bell violation against detection efficiencies of the PNRD used for photon subtraction (see Fig.~\ref{fig:bell_violation} b).}

%
%

\begin{figure}[h]
\begin{center}
\includegraphics[width=0.24\textwidth]{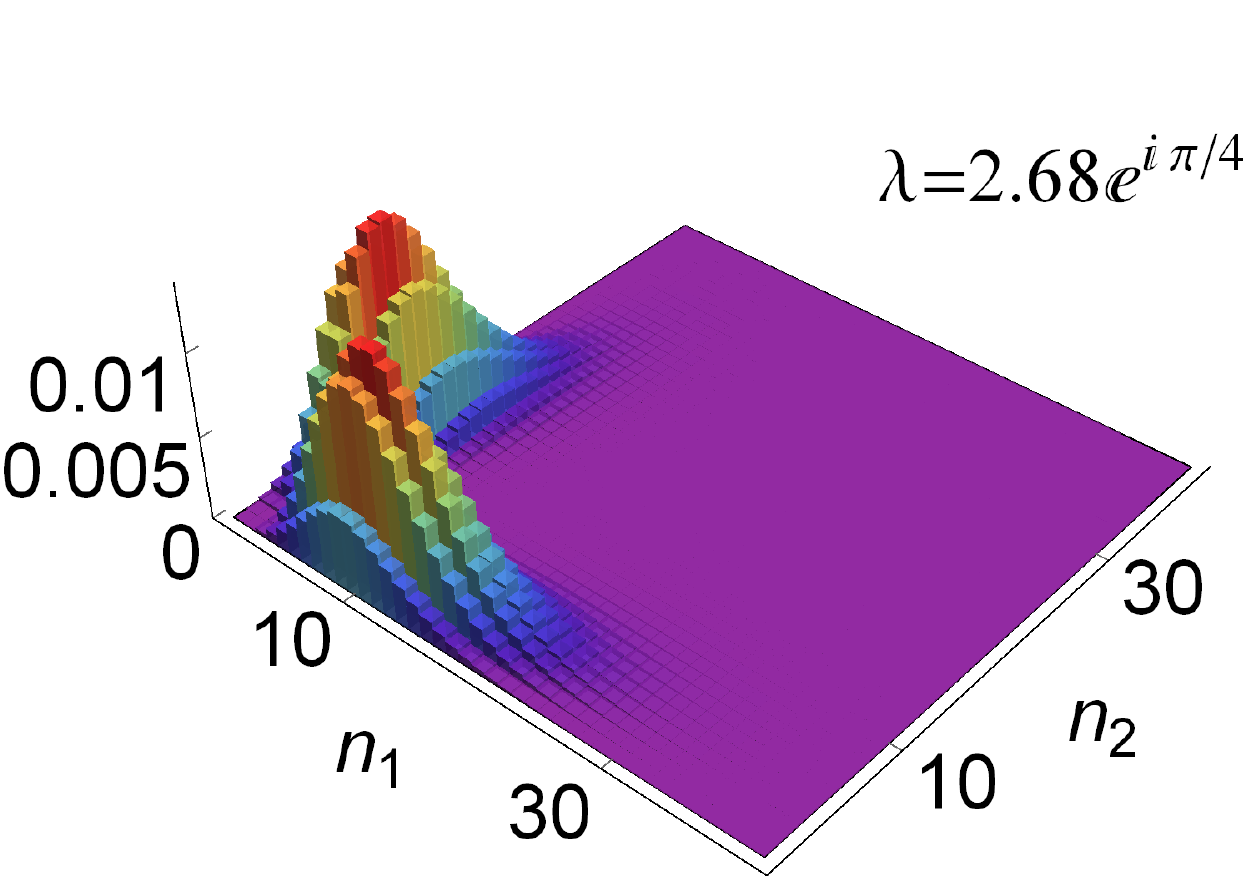}\;\;\includegraphics[width=0.24\textwidth]{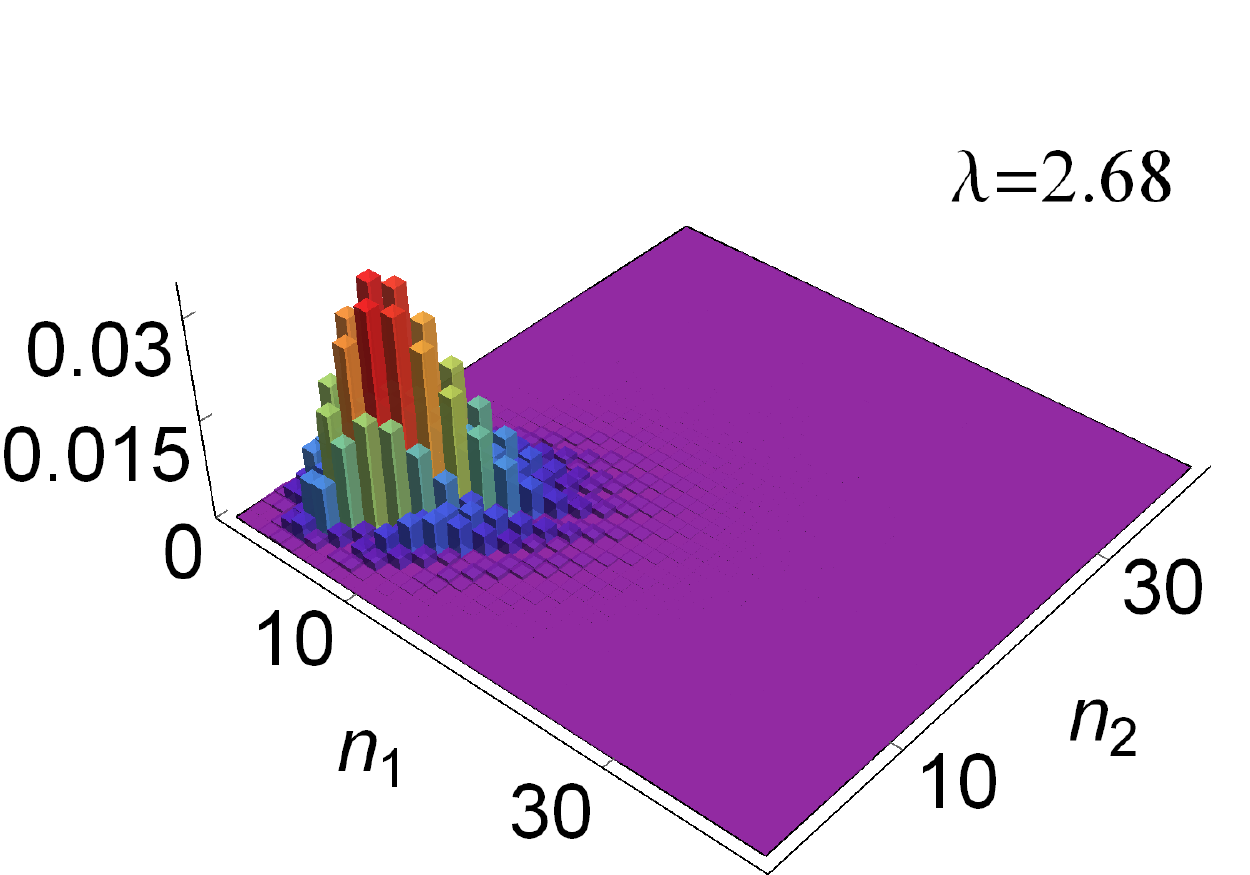}
\caption{\label{fig:pn_dist_phase}Photon number distributions after superposing a photon-subtracted squeezed vacuum state ($\zeta=1$) and a coherent state ($|\lambda|=2.68$) with non-ideal phase on a 50/50 beam splitter using the setup shown in Fig.~2 with $\tau=0.9$ and heralding on $m=3$ clicks at an ideal PNRD.}
\end{center}
\end{figure}

As a second imperfection, we consider in Fig.~\ref{fig:pn_dist_phase} the photon number distribution for an improper phase of the impinging coherent state $\ket{\lambda}_2$. {The influence of $\arg(\lambda)$ on the Bell violation is shown in Fig.~\ref{fig:bell_improper_phase}.}.

%
%

\begin{figure}[h]
\begin{center}
\includegraphics[width=0.4\textwidth]{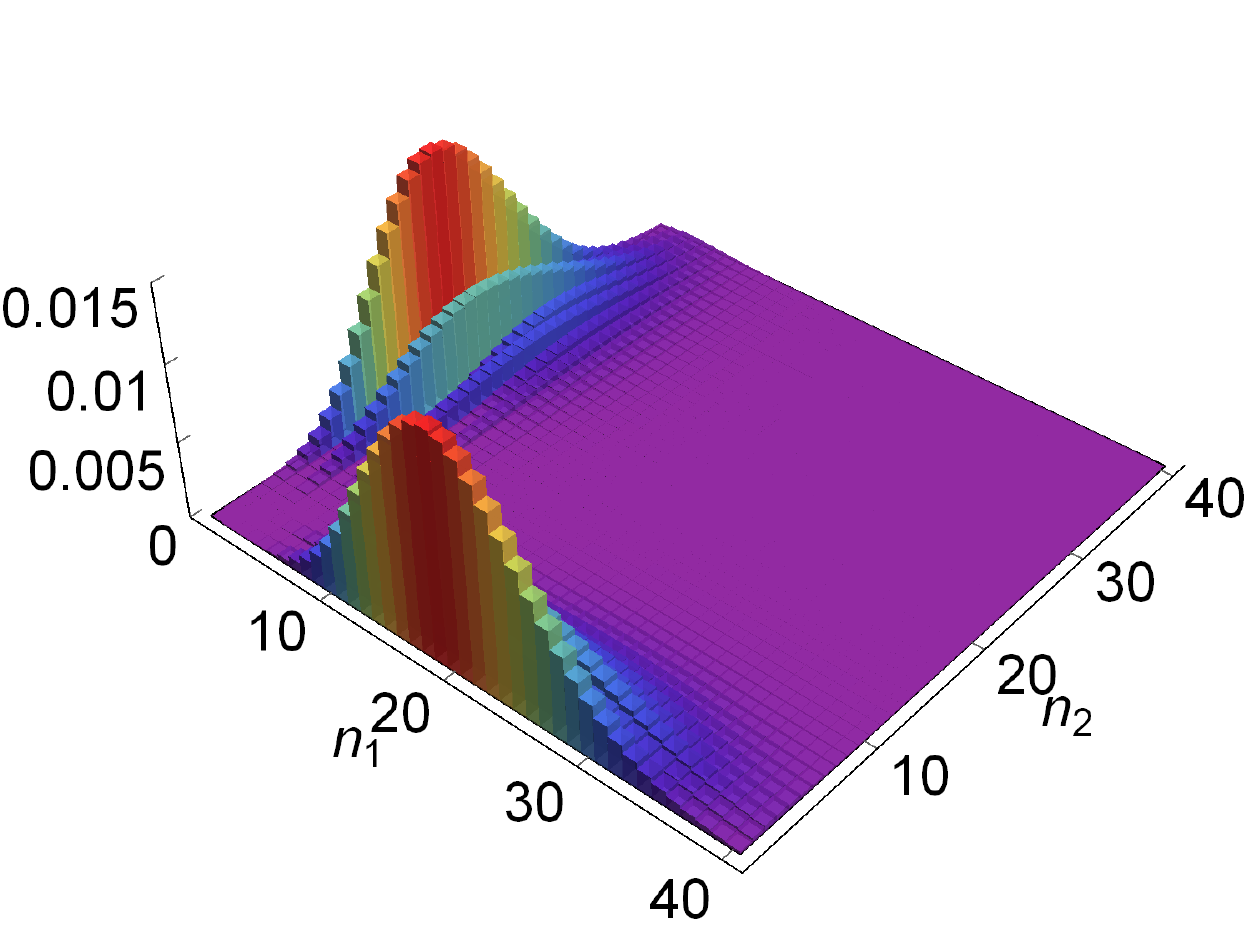}
\caption{\label{fig:pn_dist_2mode}Photon number distributions assuming a photon-subtracted two-mode squeezed vacuum ($m=3$, $\zeta=1$) and a single-mode coherent state ($\lambda=3.6\ii$) in the input port of the 50/50 beam splitters.}
\end{center}
\end{figure}

The last imperfection in the state preparation that we consider concerns multi-mode input states. Let us assume that not a single-mode squeezed state impinges on the first beam splitter but two modes each occupied with a squeezed vacuum state. As single-mode coherent states can easily be produced, we assume that in the second input port of the 50/50 beam splitter one of the two modes is occupied by a coherent state and the other by a vacuum state.  {The photon-number distribution for this situation is plotted in Fig.~\ref{fig:pn_dist_2mode}.}

%
%

\begin{figure}[h]
\begin{center}
\includegraphics[width=0.24\textwidth]{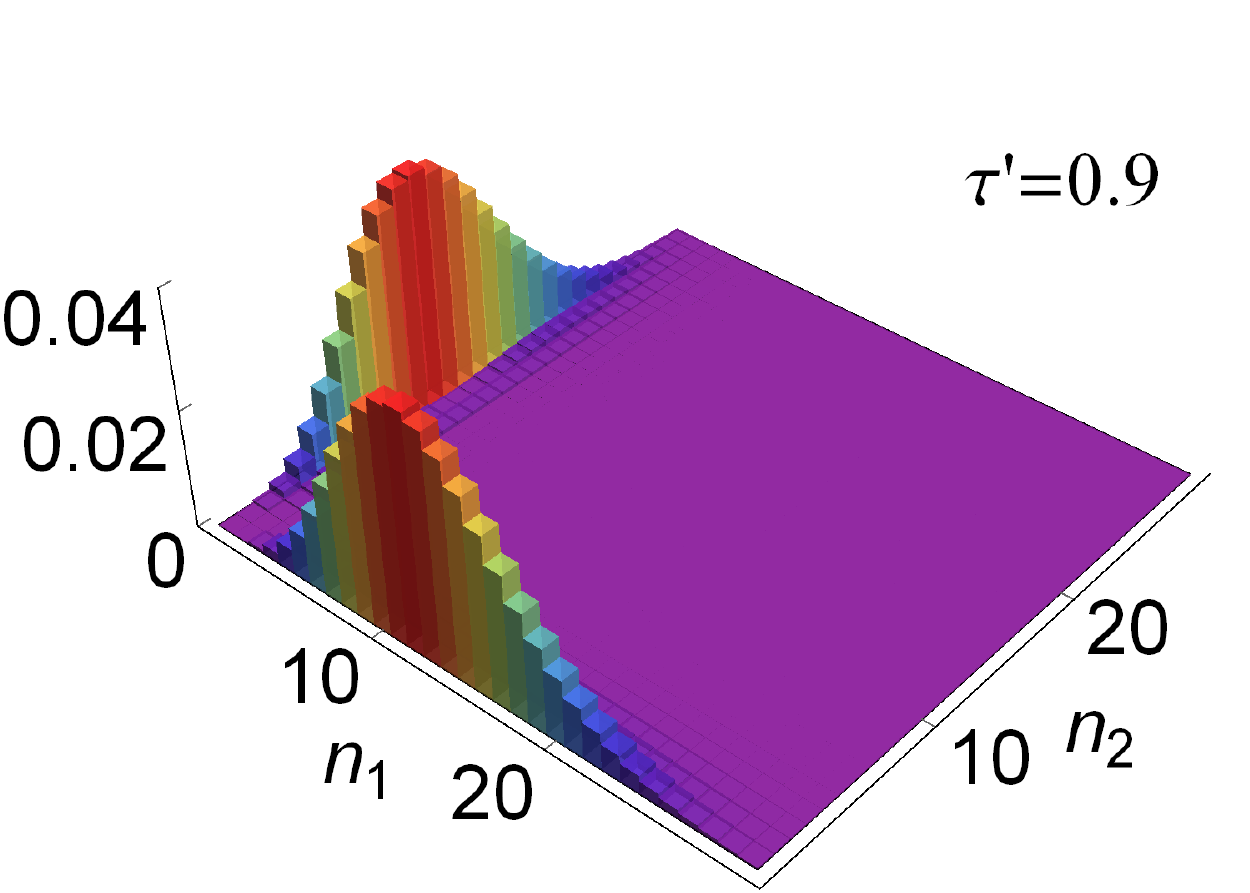}\;\;\includegraphics[width=0.24\textwidth]{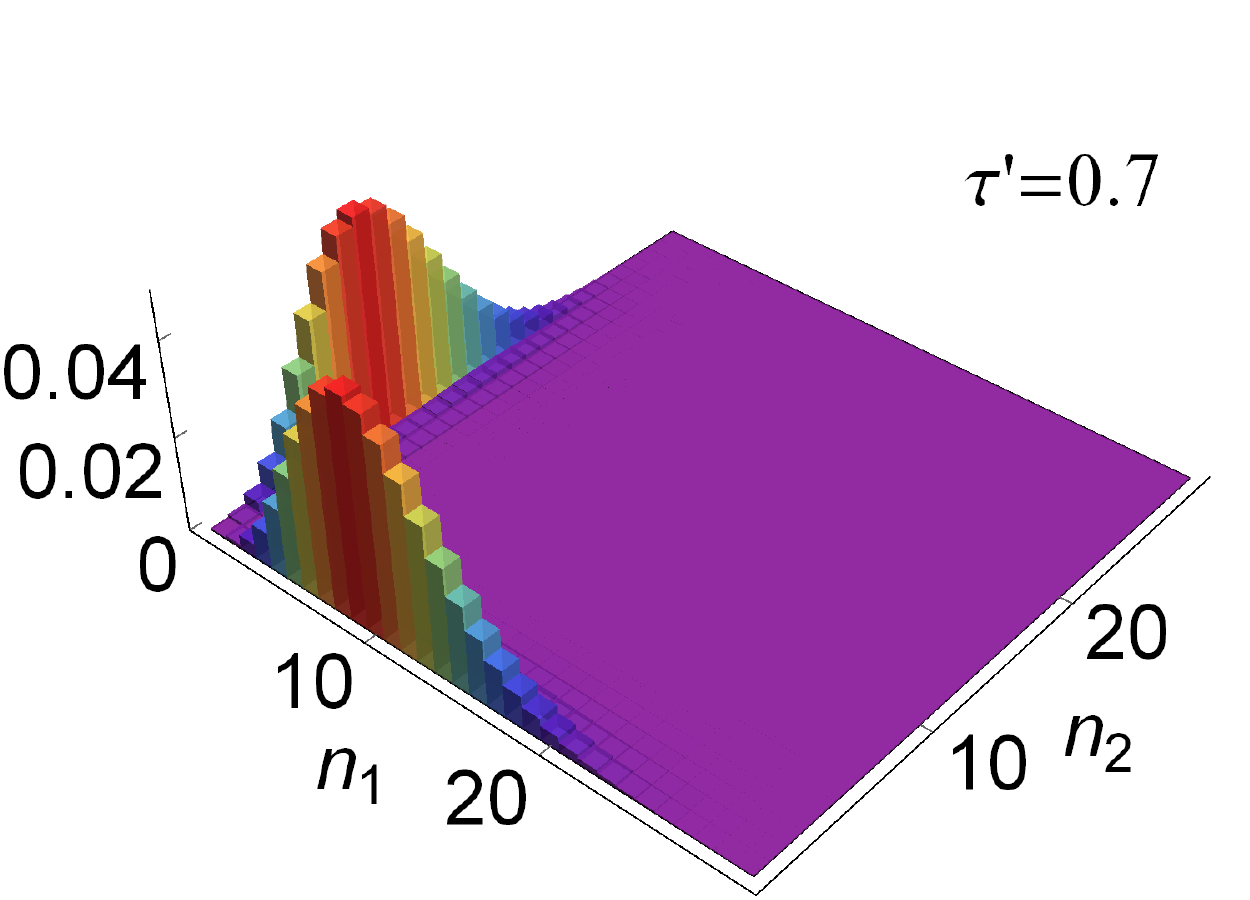}
\includegraphics[width=0.24\textwidth]{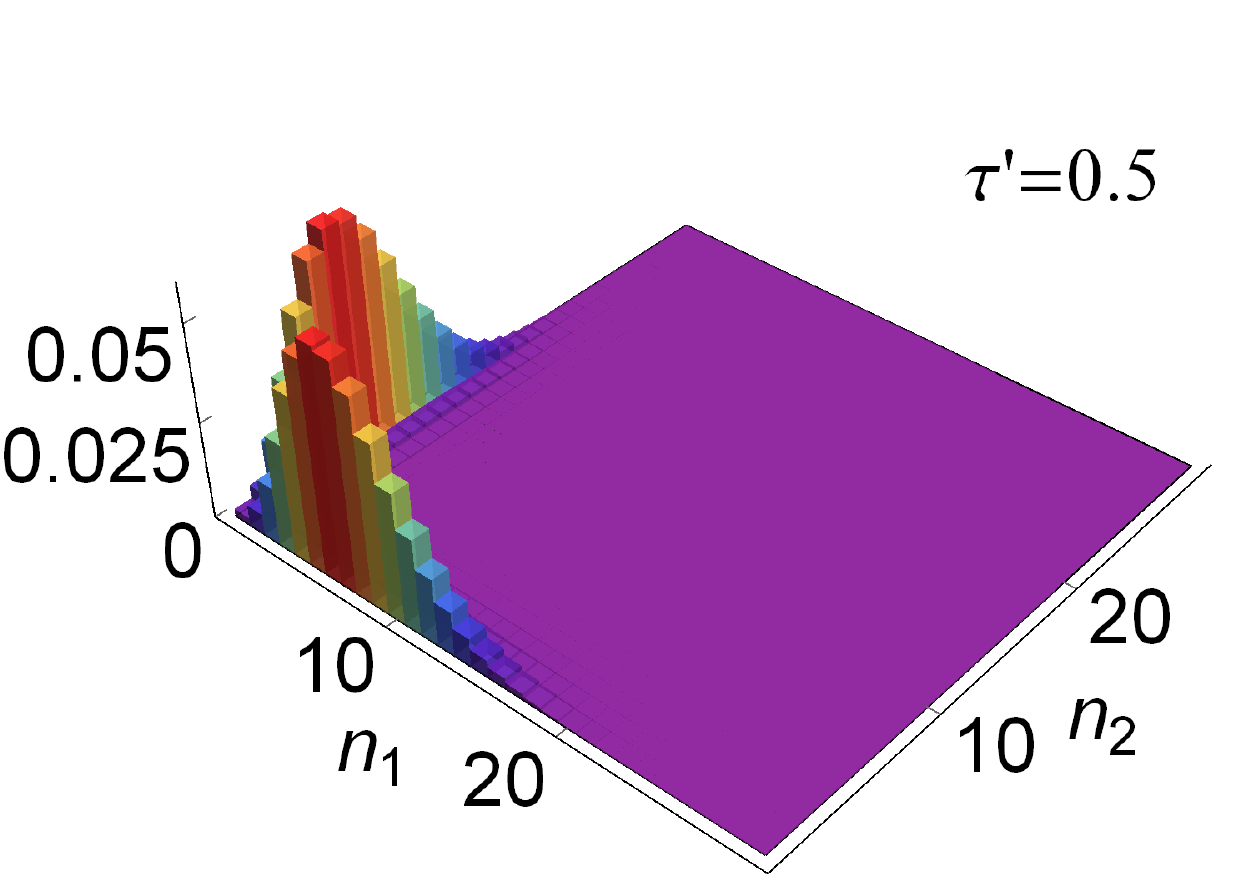}\;\;\includegraphics[width=0.24\textwidth]{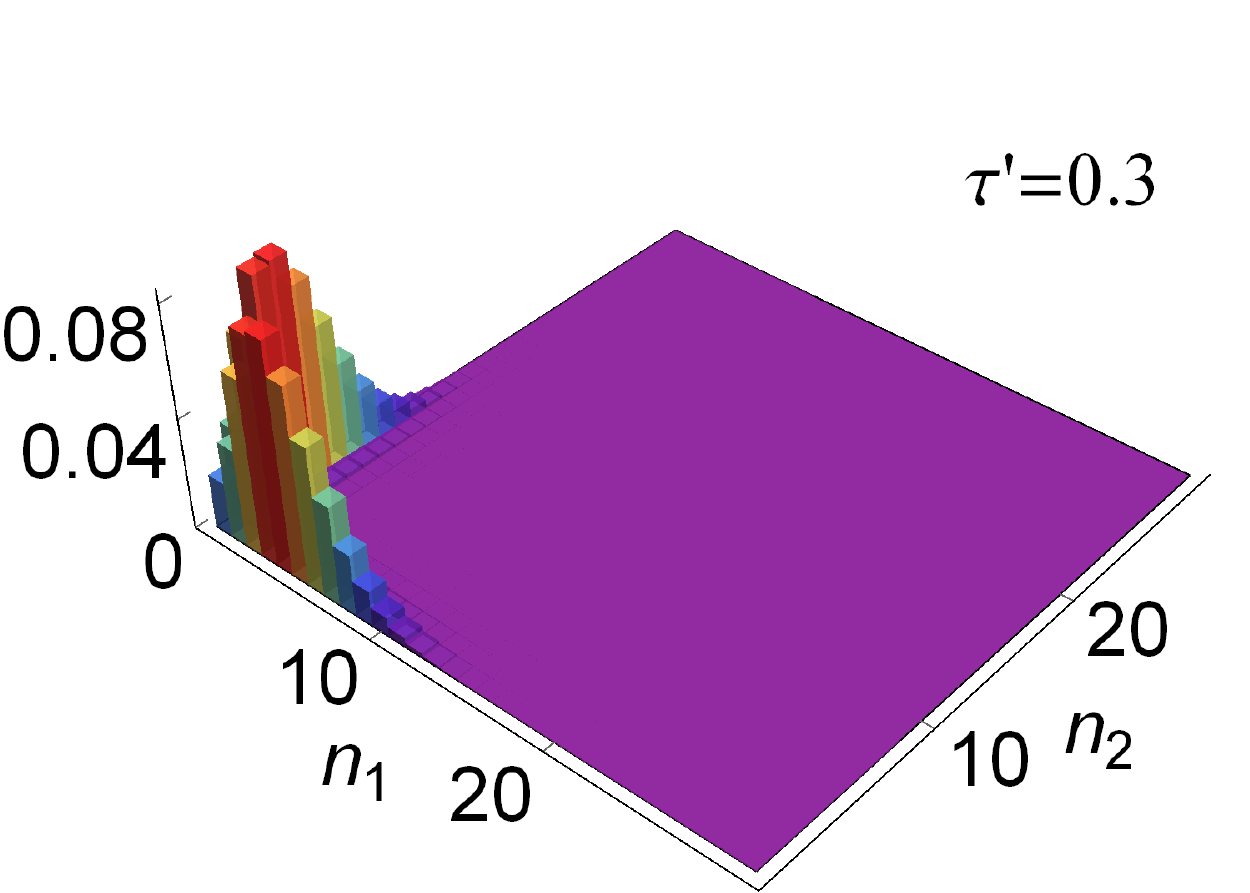}
\caption{\label{fig:pn_dist_transmission}Photon number distribution of the quantum state $\ket{\Psi_m}$ for $\zeta=1$, $\lambda=2.68\ii$ and $m=3$ when it suffers from transmission losses in the output modes. Here $\tau'$ denotes the transmittivity of the output modes 1 and 2.}
\end{center}
\end{figure}

Finally, we show in Fig. \ref{fig:pn_dist_transmission} how the photon number distribution is affected by transmission losses in the output modes. Here as well, the N00N-like structure of the output state survives in the presence of losses. {This feature is also reflected in the remarkable immunity of the Bell violation against a lossy transmission (see Fig.~\ref{fig:bell_violation} c).}

%
%

\newpage

\bibliography{bib_articles}

\end{document}